# Extending the UXR Point of View Playbook: Triangulating Insights in Complex Developer Domains

Sarah Kianfar
Google
Bellevue, WA, USA
kianfar@google.com

# ABSTRACT

As User Experience Research (UXR) matures, practitioners face the challenge of moving beyond data collection toward establishing a compelling Point of View (POV) that drives strategic impact. This paper proposes an extension to the UXR POV Playbook, specifically focusing on the transition from the "Insight Generation" layer to the "POV" layer. Drawing on extensive multi-method research in Cloud Developer Tools, spanning AI Agents, Command Line Interfaces (CLI), and Error Messages, we demonstrate how triangulating qualitative and quantitative data facilitates the creation of high-confidence POVs. We introduce three new "Playbook Cards" derived from this research: *The Paradigm Shift*, *Explainability as Trust*, and *The Cost of Friction*. These cards provide a structured mechanism for researchers to translate complex technical findings into irrefutable business narratives.

# CCS CONCEPTS

- Human-centered computing;
- Human computer interaction (HCI);
- User studies;

# KEYWORDS

# 1 INTRODUCTION

The UXR POV Playbook establishes a pyramid framework (Foundation, Data Collection, Insight Generation, and POV), to guide researchers in translating raw data into wisdom that directs business strategy [1]. However, in complex technical domains such as Artificial Intelligence (AI) and developer tooling, the leap from "Insight" to "POV" is often hindered by the difficulty of synthesizing heterogeneous data streams.

This paper presents a case study on how rigorous triangulation of qualitative and quantitative methods can bridge this gap. We analyze three distinct research initiatives to identify mechanisms that elevate research findings into strategic POVs:

- **AI Agents ("The Paradigm Shift"):** Triggered by the launch of *Gemini Code Assist Agent*, this research aimed to define the product direction. It successfully pivoted leadership's vision from an autonomous "delegate-and-forget" model to a synchronous, human-in-the-loop collaborative teammate.

- **CLI Modernization ("Explainability as Trust"):** Addressing a critical gap in UX resources for non-GUI interfaces, this initiative equipped practitioners with a playbook to tackle legacy complexity. Given that the CLI is a popular interface on Google Cloud Platform, establishing an intuitive structure was paramount.
- **Actionable Error Messages ("The Cost of Friction"):** Initiated in response to high support ticket volumes, this initiative correlated specific error codes (e.g., 403, 404) for particular services with financial costs. The goal was to operationalize user frustration to prioritize usability fixes over new features.

We propose extending the existing playbook with three specific cards derived from these initiatives. Our approach draws on the SPACE framework [2], which emphasizes that developer productivity cannot be reduced to a single metric but requires capturing satisfaction, performance, activity, communication, and efficiency.

# 2 METHODOLOGY: TRIANGULATION AS A FOUNDATION

To build a POV robust enough to influence engineering roadmaps, our research employed a "mixed-methods triangulation" strategy, aligning with the Playbook's requirement for data credibility and confidence [1].

## 2.1 AI Agents: Longitudinal Mixed-Methods

Our "Agents" research utilized a multi-phased approach, triggered by the launch of *Gemini Code Assist* Agent. The primary goal was to define the product direction in the face of conflicting visions: while leadership assumed an autonomous "fire-and-forget" model (delegate a task and return hours later), our research revealed that users desired a synchronous "collaborative teammate." We began with longitudinal Foundational Rapid Iterative Testing and Evaluation (FRITE) sessions, which used a combination of cognitive walkthroughs and interviews, with a total of 32 participants to understand user expectations and evaluate interaction models. We then conducted a survey of 134 developers to quantify learnings at scale and prioritize capabilities we had discovered in the FRITE studies. We synthesized the qualitative and quantitative data from both methods, putting bar charts next to qualitative themes and quotes, to show the full story of what users expect and why. This triangulation allowed us to successfully pivot the product strategy from an asynchronous agent to a human-in-the-loop system where users could course-correct and verify actions. This aligns with emerging frameworks in AI-assisted programming that prioritize exploration and collaboration over simple acceleration [4].

## 2.2 CLI and Errors: Heuristics and Logs

For the Command Line Interface (CLI) and Error Messages, we developed heuristics based on qualitative learnings from interviews and surveys, and quantitative data from log analysis of command usage and errors. Lack of an intuitive structure for CLI commands, and output (including error messages) being difficult to understand were among the top user challenges found in qualitative studies. The CLI UX Playbook initiative was triggered by these findings to address a gap in resources for UXers designing non-GUI interfaces. The goal was to equip UX practitioners with a playbook to improve learnability and structure of CLI interfaces. Simultaneously, the "Actionable Error Messages" initiative was driven by a high volume of support tickets related to unactionable errors. By analyzing log data on high-volume API errors (e.g., 403, 404) and combining them with support ticket volume, we identified top errors that impacted both user satisfaction and cost of support. This quantitative analysis was bolstered by the CLI UX Playbook, which emphasizes human-readable flags and actionable error recovery heuristics. Our learnings aligned with external findings that poor error message readability significantly hinders developer performance [5]. Since we linked usability (unactionable error messages) directly to financial metrics (support tickets costs), we were able to get "Stakeholder Buy-in" to make error messages more actionable [1].

# 3 EXTENDING THE PLAYBOOK: FROM INSIGHT TO POV

The transition from Insight Generation to POV requires interpreting the significance of data to articulate user needs and behaviors [1]. Based on our findings, we propose three new playbook cards to facilitate this transition. To align with the established Playbook visual style [7, 8], we present these as dual-sided cards.

## 3.1 Playbook Card: The Paradigm Shift

*Context:* Standard research often assumes the user’s relationship with software is static. However, our Agents research (FRITE sessions with 32 participants) revealed a fundamental alteration in the user-tool relationship, shifting from a passive tool to a "collaborative teammate" [4]. For example, we found that senior developers want to delegate tasks to an agent (similar to how they delegate to a junior developer) and junior developers want to brainstorm with the agents (like a code pairing buddy). This evolution necessitates a shared mental model where both user and system understand each other's capabilities [3].

| FRONT: THE CONCEPT | BACK: BEST PRACTICE |
| --- | --- |
| ⚠️ **ISSUE:** Traditional "user need" models fail to capture the dynamic expectations users have for AI agents.<br>🃏 **TYPE:** Foundation / Mental Model<br><br>**"From Tool to Teammate"**<br><br>🔗 **RELATED CARDS:** User Needs, Stakeholder Buy-in | **Why it matters:**<br>Users have distinct mental models for agents (e.g., "Junior Dev" vs. "Pairing Buddy"). Framing insights solely around utility misses the critical dimensions of trust and delegation.<br><br>**How to implement:**<br>Reframe the product strategy around *relationship dynamics*. Define the "social contract" between user and system: What |
| | is the level of autonomy? How is supervision handled?<br><br>**Application:**<br>Use this card during early concepting and PRD writing to challenge stakeholders who view the agent as a simple "chatbot." Force the definition of a collaborative design pattern (e.g., agent asks clarifying questions) rather than just an answering machine. |

## 3.2 Playbook Card: Explainability as Trust

*Context:* The "Insight Generation" layer focuses on identifying the "why" behind behavior [1]. In both our Agents and CLI research, we found that "black-box" operations were the primary blocker to adoption. Understanding agent decision making, and visibility into actions agents take were among the top capabilities requested in the survey of 134 developers, and also an emerging theme in FRITE studies. Interviews and surveys related to CLI research showed that unclear/unactionable output hinders the ability to effectively use the CLI. Research indicates that Explainable AI (XAI) fosters trust and accountability by making decision-making visible [3, 6].

| **FRONT: THE CONCEPT** | **BACK: BEST PRACTICE** |
|---|---|
| ⚠️ **ISSUE:** Users reject autonomous actions they cannot understand or verify.<br>🃏 **TYPE:** Insight Generation<br><br>**"Trust is a Function of Visibility"**<br><br>🔗 **RELATED CARDS:** Interpretation, Significance | **Why it matters:**<br>Explainability is not just a feature; it is a requirement for adoption. Users demand to know *why* an agent took action before they will trust it with future tasks.<br><br>**How to implement:**<br>Mandate that any POV regarding autonomous systems explicitly addresses how the system exposes its reasoning. "Traceable Reasoning" must be a core design principle.<br><br>**Application:**<br>Deploy this as a heuristic checklist in Design Reviews (e.g., "Does the agent show why it chose this code?"). Use it to push back on engineering requests to hide logs, establishing observability as a non-negotiable product requirement. |

### 3.3 Playbook Card: The Cost of Friction

*Context:* The POV Playbook includes a "Stakeholder Buy-in" play with cards like "Demonstrate Impact" [1]. Our research into Error Messages offers a concrete method to execute this play by mapping usability issues to financial metrics and translating findings into business narratives. Actionable error messages got leadership buy-in and funding because we showed that unclear/unactionable errors not only make user unhappy (interview/survey data), but also account for 8% of support tickets, and the cost can increase if we do nothing about it. Our triangulation and translation of user issues to business loss led to resources to improve error messages so they are actionable on their own.

| FRONT: THE CONCEPT | BACK: BEST PRACTICE |
| --- | --- |
| ⚠️ **ISSUE:** Usability issues are often de-prioritized as mere "polish" rather than critical blockers.<br>🃏 **TYPE:** Stakeholder Buy-in (Offensive Play)<br><br>**"Friction Has a Price Tag"**<br><br>🔗 **RELATED CARDS:** Demonstrate Impact, Business Awareness | **Why it matters:**<br>Quantifying friction transforms qualitative frustration into business urgency. (e.g., Unclear errors = 8% of support tickets).<br><br>**How to implement:**<br>Map usability issues directly to financial metrics. Correlate log data (e.g., 403 errors) with support ticket volume to calculate the dollar cost of poor UX.<br><br>**Application:**<br>Use this as an "Offensive Play" to secure resources. Present a "bill" to leadership (e.g., "This error costs us $50k/month") to force the prioritization of engineering tickets over new feature work. |

# 4 DISCUSSION

The UXR POV Playbook emphasizes that a POV must be "based on data, evidence, and insights" [1]. Our work demonstrates that rigorous triangulation is the engine that powers this process. By combining the broad reach of surveys with the depth of FRITE sessions and the hard truth of log analysis, researchers can construct POVs that are resilient to stakeholder bias. The proposed cards, The *Paradigm Shift, Explainability as Trust,* and The *Cost of Friction*, provide actionable frameworks for researchers to elevate their insights into strategic narratives that define the future of product development.

The impact of applying these cards has been significant across our organization. The "Cost of Friction" card led to enabling a framework for more actionable error messages that was adopted by 40+ major first-party services across Google Cloud, impacting approximately 21.6 billion errors per day (measured through internal dashboards). The "Paradigm Shift" card directly informed a VP-level product strategy change, pivoting the roadmap from a long-running background agent to a real-time chat agent with a tight human and agent feedback loop leading to the launch of Gemini Code Assist "Agent Mode" preview. Finally, the "Explainability as Trust" card, codified in the CLI UX Playbook, has been adopted by several divisions and shaped the interaction design of Gemini CLI.

# 5 ACKNOWLEDGEMENTS

The various research projects mentioned in this paper would not have been possible without the amazing cross-functional collaboration with my colleagues. I would like to express my deepest gratitude to *Prithpal Bhogill* and *Luna Wang* for the AI Agents project; *Santosh Mathan*, *Miguel Solorio*, and *Josh Hirshfeld* for the CLI UX Playbook; and *Jorge Cueto* and *Bryan Zimmerman* for the Actionable error messages project.

# REFERENCES

1. Stephen Giff, Renée Barsoum, and Huseyin Dogan. 2024. User Experience Research: Point of View Playbook. In *Extended Abstracts of the CHI Conference on Human Factors in Computing Systems (CHI EA '24)*. ACM.
2. Forsgren, N., Storey, M. A., Maddila, C., Zimmermann, T., Houck, B., & Butler, J. (2021). The SPACE of Developer Productivity: There's more to it than you think. *Queue*, 19(1), 20-48.
3. Q. Vera Liao and S. Shyam Sundar. 2022. Designing for Responsible Trust in AI Systems: A Communication Perspective. In *Proceedings of the 2022 CHI Conference on Human Factors in Computing Systems (CHI '22)*. ACM.
4. Advait Sarkar. 2023. Generalized, constant-time, and collaborative: The future of AI-assisted programming. *IEEE Software* 40, 4 (2023), 20–26.
5. Titus Barik, Justin Smith, Kevin Lubick, Elisabeth Holmes, Jing Feng, Emerson Murphy-Hill, and Chris Parnin. 2017. Do developers read compiler error messages?. In *Proceedings of the 39th International Conference on Software Engineering (ICSE '17)*. IEEE Press.
6. Naiseh, M., Dogan, H., Giff, S., & Jiang, N. (2025). Development of a persuasive User Experience Research (UXR) Point of View for Explainable Artificial Intelligence (XAI). In *CHI 2025 Workshop on UXR POV*.
7. Lau, J. & Tran, A. (2025). UXR Point of View on Product Feature Prioritization Prior To Multi-Million Engineering Commitments. In *CHI 2025 Workshop on UXR POV*.
8. Yew, J., Lee, D., & Du, R. (2025). Developer productivity metrics: a developer experience focused perspective. In *CHI 2025 Workshop on UXR POV*.